\documentclass[prd,floatfix,nofootinbib,showpacs,showkeys]{revtex4}
\usepackage{exscale}                  
\usepackage[intlimits]{amsmath}       
\usepackage{amsfonts}
\usepackage{amssymb,amscd}
\usepackage[dvips]{epsfig}                   
\usepackage{array}
\usepackage{wrapfig}
\newcommand{\beqa}{\begin{eqnarray}}
\newcommand{\eeqa}{\end{eqnarray}}
\newcommand{\beq}{\begin{equation}}
\newcommand{\eeq}{\end{equation}}
\newcommand{\gS}[1]{#1\!\!\!\!\!\not~}	
\newcommand{\GS}[1]{#1\!\!\!\!\!\!\!\not~}	
\newcommand{\Dslash}{\GS{D}}
\newcommand{\qslash}{\gS{q}}
\newcommand{\pslash}{\gS{p}}
\newcommand{\Pslash}{\GS{P}}

\begin{document}

\large
\hfill\vbox{\hbox{DCPT/05/136}
              \hbox{IPPP/05/68}}
\vspace{2mm}

\title{Finite volume effects in a quenched lattice-QCD quark propagator}
\author{C.~S.~Fischer,
}
\author{M.~R.~Pennington}
\affiliation{IPPP, Durham University, Durham DH1 3LE, U.K. }
\date{\today}

\begin{abstract}
We investigate finite volume effects in the pattern of chiral 
symmetry breaking. To this end we employ a formulation of the
Schwinger-Dyson equations on a torus which reproduces results 
from the corresponding lattice simulations of staggered quarks and 
from the overlap action. Studying the volume dependence of the
quark propagator we find quantitative differences with the infinite 
volume result at small momenta and small quark masses. We estimate 
the minimal box length $L$ below which chiral perturbation theory 
cannot be applied to be $L \simeq 1.6$ fm. In the infinite volume limit 
we find a chiral condensate of 
$|\langle \bar{q}q\rangle|_{\overline{MS}}^{2 GeV} = (253 \pm 5 \,\,\mbox{MeV})^3$, 
an up/down quark mass of 
$m_{\overline{MS}}^{2 GeV} = 4.1 \pm 0.3 \,\, \mbox{MeV}$
and a pion decay constant which is only ten percent smaller than 
the experimental value. 
\end{abstract}


\maketitle


\section{Introduction}\label{sec:intro}

Dynamical chiral symmetry breaking is certainly one of the most interesting 
phenomena of QCD. It is entirely a strong coupling effect in the sense that
dynamical quark masses cannot be generated at any order in perturbation theory.
Thus nonperturbative methods like lattice Monte-Carlo simulations
\cite{Chandrasekharan:2004cn}, chiral perturbation theory \cite{Leutwyler:2000jg}
or the Green's function approach using the Schwinger-Dyson 
and Bethe-Salpeter equations (SDE/BSE) \cite{Alkofer:2000wg,Maris:2003vk},
are needed to explore chiral symmetry and its breaking pattern. 

With the rediscovery of the Ginsparg-Wilson relation and the construction 
of actions for overlap, domain wall and perfect fermions, the lattice 
formulation of QCD emerged in principle as an appropriate nonperturbative 
tool, with which to study the effects of dynamical chiral symmetry breaking. 
However in practice, lattice simulations with small quark masses are extremely 
expensive in terms of CPU-time. It is only with staggered fermion actions that 
quark masses not far from their physical values have been achieved to date, but these 
actions have the disadvantage that full chiral symmetry is only recovered in the 
continuum limit. Consequently, there is no certainty  with any finite volume that 
the correct breaking pattern can be observed.

Lattice simulations are, of course, always performed at a finite volume. Since continuous
symmetries cannot be spontaneously broken at a finite volume $V$, chiral symmetry 
is restored in the limit of zero quark mass, $m \rightarrow 0$, independently of
the formulation of the lattice action. Thus one first has to perform the limit
$V\rightarrow \infty$ before one can investigate the chiral limit. At the hadron level, 
whether for mesons and baryons, chiral perturbation theory provides a reliable tool with 
which to make such an extrapolation (see e.g. \cite{Detmold:2001jb,Procura:2003ig,Colangelo:2005gd} 
and references therein). Volume effects for any particles that couple to the pion
can be arranged in powers of $\exp[-M_\pi L]$, where $M_\pi$ is the pion mass
and $L$ is the size of the box \cite{Gasser:1987zq}. On the other hand, chiral 
perturbation theory has 
nothing to say about volume effects in the underlying quark and gluon substructure. 
For this the Green's function approach employing Schwinger-Dyson equations provides 
a suitable alternative. Indeed, a recent investigation of the gluon and ghost 
propagator of the Landau gauge Yang-Mills theory on a torus could offer an explanation 
for systematic differences in the infrared behaviour of these propagators in a 
box, compared with the infinite volume limit \cite{Fischer:2005ui}.

In the present work we extend this analysis and investigate finite volume effects 
in the quark propagator. To this end we study the QCD-gap equation for
the quark propagator on a torus. The input from the Yang-Mills sector of QCD 
consists of numerical solutions for the ghost and gluon propagator, which match 
corresponding lattice calculations \cite{Bowman:2004jm,Sternbeck:2005tk}.
Presently unknown contributions from the quark-gluon vertex are parametrised such 
that the quenched lattice quark propagators from each of Ref.~\cite{Bowman:2002bm} (staggered)
and Ref.~\cite{Zhang:2004gv} (overlap) are reproduced by the gap equation. This idea has 
already been explored to some extent by Bhagwat {\it et al.}~\cite{Bhagwat:2003vw}. 

Our treatment differs from that of Ref.~\cite{Bhagwat:2003vw} in two essential respects: 
firstly, we solve the coupled set of three SDEs for the ghost, gluon and quark 
propagators, and so include the Yang-Mills sector of the SDEs in the Green's function 
approach. Secondly, we calculate our propagators on a manifold similar to that of the lattice 
and fit the interaction to reproduce the lattice results of the gluon and quark propagators 
on their respective manifolds. 

One of the advantages of the Green's function approach is that volume effects
can be studied continuously from very small to very large volumes
(corresponding studies for meson observables using chiral perturbation 
theory {\it e.g.} have to distinguish between two different regions of chiral 
counting \cite{Colangelo:2005gd}).
Furthermore one has direct access to the infinite volume and the continuum limit without
the need to perform any extrapolations. We are thus in a position to study
chiral symmetry restoration at small volumes together with effects at large and infinite
volumes in the same framework.

The paper is organised as follows: In the next section we shortly review basic properties of
the pattern of dynamical chiral symmetry breaking in a box. In particular we recall the 
derivation of the Casher-Banks relation and a basic estimate for a minimal box length for
chiral perturbation theory. In section \ref{sec:dse} we discuss the technical details 
associated with solving the quark Schwinger-Dyson equation on a torus. Our results on
the compact manifold are 
presented in section \ref{sec:Numres}. We start with a summary of finite volume effects 
in the Yang-Mills sector based on the results of Ref.~\cite{Fischer:2005ui}. These are 
used as input into the quark-SDE together with a model for the quark-gluon interaction, 
which is discussed in section \ref{sec:YM}. The interaction is fitted so that the  
lattice data for the quark propagator from Ref.~\cite{Bowman:2002bm} (staggered) and 
Ref.~\cite{Zhang:2004gv} (overlap) respectively are reproduced by the quark-SDE. We then determine
the corresponding quark propagator at larger volumes and in the infinite volume
limit. The results are compared in section \ref{sec:quark}. For small quark masses we 
find sizeable quantitative volume effects which are still present at comparably large volumes.
These effects are qualitatively similar for both the propagators from staggered and overlap
quarks.
The small volume behaviour of the quark propagator is investigated in section \ref{sec:XPT}.
For a fixed, small current quark mass we determine the onset of dynamical chiral symmetry 
breaking when the volume of the box is increased. We find a minimal box length of
$L \simeq 1.6$ fm, below which chiral perturbation theory cannot be safely applied.
Finally, we discuss the infinite volume properties of our quark propagator. 
We determine the chiral condensate in section \ref{sec:condensate}, comment
on possible analytic structures in section \ref{sec:analytical} and give results
for the corresponding pion mass and decay constant in section \ref{sec:pion}. We 
summarise and conclude in section \ref{sec:sum}.

\section{Chiral symmetry breaking in a box}\label{sec:chiral}

Before we embark on our investigation, let us recall the finite volume behaviour of the 
chiral condensate, as this is
the order parameter of dynamical chiral symmetry breaking,~\cite{Leutwyler:1992yt}.
The fermion propagator in its spectral representation is given by
\beq
S_A(x,y) \;=\; \sum_n \frac{u_n(x)\, u^\dagger_n(y)}{m-i\lambda},
\eeq
where $u_n(x)$ and $\lambda_n$ are eigenfunctions and eigenvalues of the
Euclidean Dirac operator, $\Dslash u_n(x) = \lambda_n u_n(x)$. The gauge field
$A$ is treated as an external field. These 
eigenfunctions occur either as zero modes or in pairs of opposite eigenvalues.
Setting $x=y$, integrating over x and neglecting the zero mode contributions,
one obtains
\beq
\frac{1}{V} \int_V S_A(x,x) \;=\; -\frac{2m}{V} \sum_{\lambda_n > 0} 
\frac{1}{m^2+\lambda_n^2}. \label{eq2}
\eeq 
The quark condensate can be deduced by averaging  the left hand side of this equation over
all gauge field configurations and then taking the infinite volume limit to give
\beq
\langle \bar{q}q\rangle \;=\; -2m \int^\infty_0 d\lambda \;
\frac{\rho(\lambda)}{m^2 + \lambda^2}, 
\eeq
where $\rho(\lambda)$ is the mean level density of the spectrum, which becomes 
dense in the infinite volume limit. In the chiral limit, $m \rightarrow 0$, only
the infrared part of the spectrum contributes and one finally arrives at the 
Banks-Casher relation \cite{Banks:1979yr}
\beq
\langle \bar{q}q\rangle \;=\; -\pi \rho(0) \label{Banks-Casher}\; .
\eeq

\noindent If the two limits are interchanged, {\it i.e.} if one takes the chiral limit before
the infinite volume limit, one has a discrete sum in Eq.~(\ref{eq2}) and
the infrared part of the spectrum cannot trigger a nonvanishing chiral condensate:
chiral symmetry is restored. If, however, at a given volume the explicit quark mass
$m$ is not too small, one can still observe the spontaneous formation of a quark 
condensate. If the factor $(m^2+\lambda_n^2)^{-1}$ varies only slightly with 
$n$,
the sum in Eq.~(\ref{eq2}) can still be replaced by an integral and Eq.~(\ref{Banks-Casher})
remains valid. For this to be a legitimate approximation
one needs $m \gg \Delta \lambda \sim 1/V\rho(\lambda) = \pi/(V|\langle \bar{q}q\rangle|)$,
at the lower end of the spectrum. Thus one obtains the condition
\beq
V m |\langle \bar{q}q\rangle| \gg \pi.
\eeq
To get a feeling for this condition, note that for a typical value of the 
chiral condensate of $|\langle \bar{q}q\rangle| = (0.25 {\rm ~GeV})^3$ and a volume of
$V = (5 {\rm ~fm} )^4$ the quark mass has to be of the order $m \gg 5 \cdot 10^{-4}$ GeV, which
is well satisfied for all quark masses of physical interest. Whether there are
sizeable modifications to the corresponding quark propagator due to the box is
however a different question, as we shall see in section \ref{sec:quark}.

Chiral perturbation theory builds upon the chiral limit, {\it i.e.} it can only be applied 
on volumes large enough such that small quark masses remain accessible. Correspondingly
the chiral expansion parameter $p/(4\pi f_\pi)$ has to be small. On a torus the bosonic 
degrees of freedom have momenta ${\bf p} = 2\pi {\bf n}/L$ with ${\bf n}$ a vector of 
integers. Small nonzero momenta are therefore only present if the condition 
\beq
L >> \frac{1}{2 f_\pi} \sim 1 {\rm ~fm}
\eeq
is satisfied. {\it A priori} there is no way to say by how much $L$ has to exceed 1 fm
\cite{Colangelo:2005gd}. In section \ref{sec:XPT} we will see that this scale can be estimated 
using the quark-SDE on a torus.

\section{The quark Schwinger-Dyson equation on a torus}\label{sec:dse}

\begin{figure}[t]
\centerline{\epsfig{file=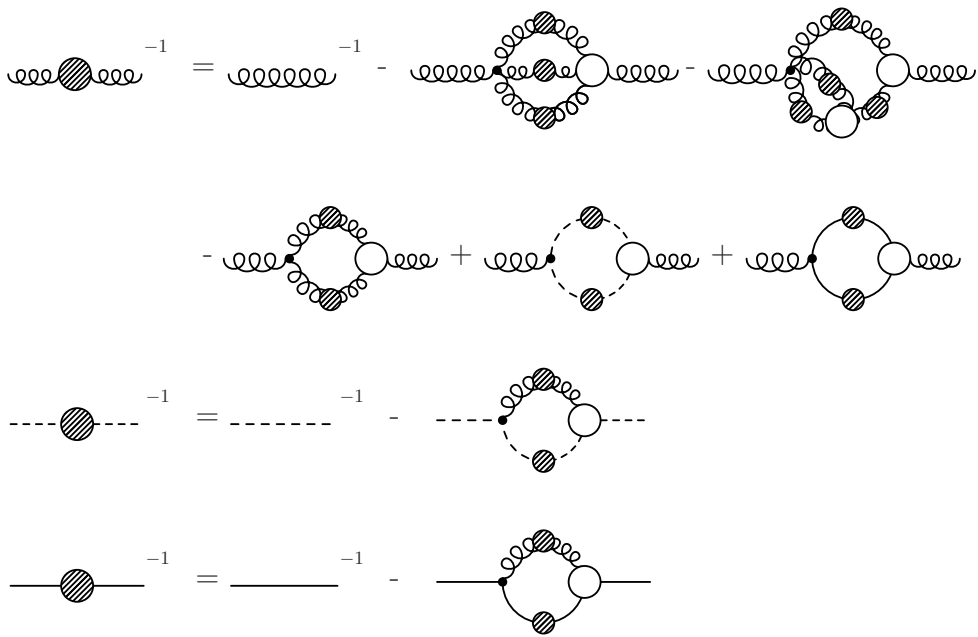,width=13cm}}
\caption{The coupled set of Schwinger-Dyson equations for the gluon, ghost
and quark propagators.}
\label{DSEs}
\vspace{5mm}
\end{figure}

In Euclidean momentum space, the renormalised dressed ghost, gluon and quark propagators in the 
Landau gauge are given by
\vspace{-2mm}
\beqa
D^G(p^2)          &=& - \frac{G(p^2)}{p^2} \,, \\[1.5mm]
D_{\mu \nu}(p)    &=&   \left(\delta_{\mu \nu} -\frac{p_\mu p_\nu}{p^2}\right) 
                        \frac{Z(p^2)}{p^2} \,, \\[1.5mm]
S(p)              &=& \frac{Z_f(p^2)}{i \pslash + M(p^2)} 
\eeqa
Here the ghost dressing function $G(p^2)$, the gluon dressing function $Z(p^2)$ and
the quark wave function renormalisation $Z_f(p^2)$ also depend  on the renormalisation 
point $\mu^2$, whereas the quark mass function $M(p^2)$ is a renormalisation group 
invariant. These propagators are given by their corresponding Schwinger-Dyson
equations shown diagrammatically in Fig.~\ref{DSEs}.  Since we aim to analyse 
a quenched lattice quark propagator we will also work in the quenched approximation 
in the SDE approach and so neglect the quark loop in the gluon-SDE. An estimate of
unquenching effects for the propagators and for light meson observables can be found in
Ref.~\cite{Fischer:2005en}.

On a compact manifold, the ghost, gluon and quark fields have to obey appropriate
boundary conditions in the time direction. These have to be periodic for
the gluon and ghost fields\footnote{The condition for the ghost field can
be read off easily from its BRST-transformation \cite{Fischer:2005ui}.}, and antiperiodic 
for the quarks. It is convenient, though not necessary, to choose the same conditions
in the spatial directions\footnote{Different conditions in the spatial directions have been
explored in \cite{Aoki:1993gi,Braun:2005gy}.}. We choose the box to be of equal length in all 
directions, $L_1=L_2=L_3=L_4\equiv L$, and denote the corresponding volume $V=L^4$.
Together with the boundary conditions this leads to discretised momenta in momentum
space. Thus all momentum integrals appearing in the Schwinger-Dyson equations are
replaced by sums over Matsubara modes. Since the ghost and gluon SDE on a torus 
have been investigated in detail in Refs.~\cite{Fischer:2005ui,Fischer:2002eq}, 
we only discuss the quark-SDE here. On the manifold $R^4$, the quark-SDE can be written as
\beq
S^{-1}(p) \;=\; Z_2 \, [S^0(p)]^{-1} -  C_F \, \frac{Z_2}{\widetilde{Z}_3}\, 
\frac{g^2}{(2\pi)^4} \int d^4{k} \,
\gamma_{\mu}\, S(k) \,\Gamma_\nu(k,p) \,D_{\mu \nu}(p-k) \,, \label{quark}
\eeq
where the factor $C_F = 4/3$ stems from the colour trace and we have introduced a reduced 
quark-gluon vertex $\Gamma_\nu(k,p)$, by defining
$\Gamma^{full}_{\nu,i}(k,p) = i g \frac{\lambda_i}{2} \Gamma_\nu(k,p)$. The bare quark 
propagator is given by $[S^0(p)]^{-1} = i \gamma \cdot p + Z_m m(\mu^2)$, where
$m(\mu^2)$ is the renormalised current quark mass. The wave function and quark mass
renormalisation factors, $Z_2$ and $Z_m$, are determined in the renormalisation process. 
The ghost renormalisation factor, $\widetilde{Z}_3$, will be discussed below, when we introduce
our expression for the quark-gluon vertex. The quark mass
function $M(p^2)$ and the wave function $Z_f(p^2)$ can be extracted from Eq.~(\ref{quark}) 
by suitable projections in Dirac-space. 

Note that the quark propagator determined from
Eq.~(\ref{quark}) is independent of the regularisation procedure. In our numerical calculations
we use a subtracted version of Eq.~(\ref{quark}) and an O(4)-invariant UV-cutoff (for details 
see e.g. ref.~\cite{Fischer:2003rp}). It is a simple matter to explicitly verify numerically 
(and also analytically) that the resulting quark propagator is independent of the cutoff, 
which therefore can be sent to infinity at the end of each calculation. The quark-SDE,
Eq.~(\ref{quark}), therefore represents not only the infinite volume limit but also the
continuum limit (in coordinate space) of any representation of the SDE on a compact manifold.
We will use the phrase infinite volume/continuum limit in the following to indicate this
simultaneous removal of both an ultraviolet and an infrared cutoff.

\begin{figure}[t]
\centerline{\epsfig{file=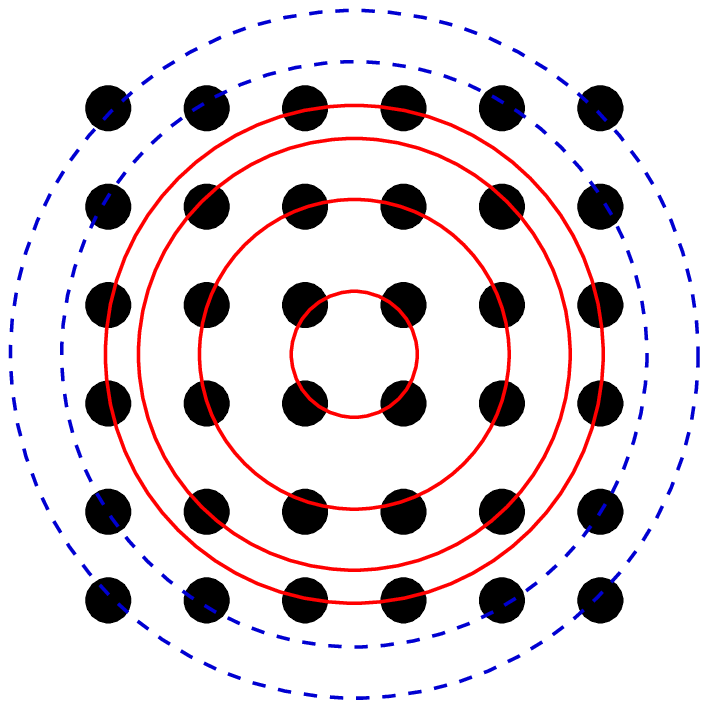,width=6cm}}
\caption{Two-dimensional sketch of the momentum grid dual to the four-torus
for a fixed Cartesian momentum cutoff. The hyperspheres depicted by dashed 
lines are not complete in the sense that additional momentum points on
these spheres are generated if the cutoff is increased. The O(4)-invariant 
cutoff used in our calculations sums only over complete hyperspheres, 
which are indicated by fully drawn circles.}
\label{fig:latt}
\vspace{5mm}
\end{figure}

On a torus with antiperiodic boundary conditions for the quark fields, the
momentum integral changes into a sum of Matsubara modes,
\beq
\int \frac{d^4q}{(2 \pi)^4} \:(\cdots) \:\:  \longrightarrow \:\:\frac{1}{L^4}
\sum_{n_1,n_2,n_3,n_4} \:(\cdots) \,, 
\eeq 
counting momenta ${\bf q}_{\bf n} = \sum_{i=1..4} (2\pi/L)(n_i+1/2) \hat{e}_i$, where
$\hat{e}_i$ are Cartesian unit vectors in Euclidean momentum space. For the numerical
treatment of the equations it is convenient to rearrange this summation 
so that it represents a spherical coordinate system \cite{Fischer:2002eq}, 
see Fig.~\ref{fig:latt} for an illustration. We then write
\vspace{2mm}  
\beq
\frac{1}{L^4} \sum_{n_1,n_2,n_3,n_4} (\cdots) \: =  \frac{1}{L^4} \sum_{j,m} \:(\cdots) \,, 
\eeq 
where $j$ counts spheres with ${\bf q_n q_n}=\textrm{const}$, and 
$m$ numbers the grid points on a given sphere. The corresponding momentum vectors are
denoted ${\bf q}_{m,j}$ and their absolute values are given by 
$q_{m,j} = |{\bf q}_{m,j}|$. It is then a simple matter to
introduce an $O(4)$-invariant cut-off by restricting $j$ to an interval [1,N].
The resulting quark SDE is given by
\beq
S^{-1}(p_{i,l}) \;=\; Z_2 \, [S^0(p_{i,l})]^{-1} -  C_F\, \frac{Z_2}{\widetilde{Z}_3}\, \frac{g^2}{L^4} \sum_{j,m}^N \,
\gamma_{\mu}\, S(k_{j,m}) \,\Gamma_\nu(k_{j,m},p_{i,l}) \,D_{\mu \nu}(p_{i,l}-k_{j,m}) \,. \label{quark_t}
\eeq
Note that the momentum argument of the gluon propagator is a difference of two
antiperiodic Matsubara momenta and thus lives on a momentum grid corresponding
to periodic boundary conditions as it should.

The quark Schwinger-Dyson equations, Eqs.~(\ref{quark}) and (\ref{quark_t}), can be solved
numerically employing well established methods once the input from the Yang-Mills sector,
the gluon propagator $D_{\mu \nu}$ and the fully dressed quark-gluon vertex 
$\Gamma_\nu(k,p)$ are specified. Our numerical method on the torus is outlined
in Ref.~\cite{Fischer:2005ui}, the corresponding continuum method as well as 
details on the renormalisation procedure of the quark-SDE are given in 
Ref.~\cite{Fischer:2003rp}. The truncation scheme of the Yang-Mills sector 
is discussed in Refs.~\cite{Fischer:2002eq}.

\section{Numerical Results for the propagators}\label{sec:Numres}

\subsection{Yang-Mills sector and parameter fitting}\label{sec:YM}

\begin{figure}[t]
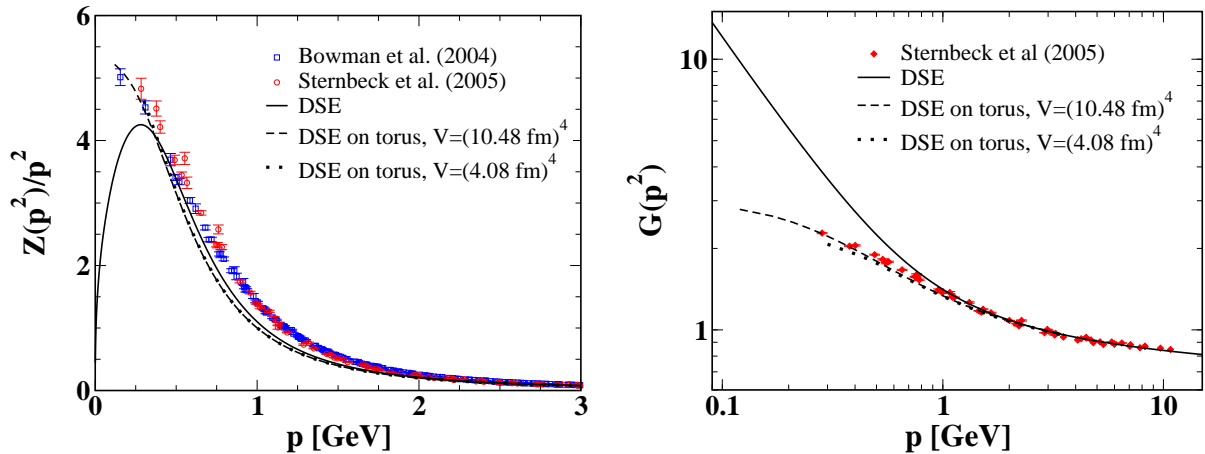

\centerline{
  \epsfig{file=torquark.glue.eps,width=7.7cm}
  \hfill
  \epsfig{file=torquark.ghost.eps,width=7.7cm}}
\caption{The results for gluon and ghost from Dyson-Schwinger equations 
in the continuum and on the torus are compared with the lattice data of
refs.~\cite{Bowman:2004jm,Sternbeck:2005tk}\label{vol2}}
\vspace{5mm}
\end{figure}

Before we discuss our ansatz for the quark-gluon vertex,
let us shortly summarise the results of Ref.~\cite{Fischer:2005ui} for the ghost and
gluon propagators on the torus. In Fig.~\ref{vol2},
the numerical solutions for the gluon propagator $Z(p^2)/p^2$ (left diagram) and the 
ghost dressing function $G(p^2)$ (right diagram) in the continuum
and on a torus\footnote{There are two slight changes compared to the
treatment in Ref.~\cite{Fischer:2005ui}: we adapted the overall scale to match the 
lattice results and the ultraviolet cutoff has been increased from 
$\Lambda=2.47$ GeV to $\Lambda=3.60$ GeV in order to minimise artifacts due to the cut-off.} 
are displayed together with the results of recent lattice simulations.
Overall there is very good agreement between the DSE-solutions on the compact manifold 
and the lattice data. However, the infrared (IR) behaviour of the DSE-solutions on $R^4$
is qualitatively different, although the truncation scheme used in solving the SDEs
is the same. The ghost dressing function in Fig.~\ref{vol2} diverges
in the infinite volume/continuum limit, whereas it stays finite on the compact manifold. 
For the gluon propagator, this difference can be expressed in terms
of an infrared power law, 
\beq
Z(p^2) \sim (p^2)^{2\kappa}.\label{Z_IR} 
\eeq
One obtains
$\kappa \approx 0.5$ (IR-finite) on a compact manifold (even for very large volumes), 
whereas $\kappa \approx 0.596$ (IR-vanishing) on $R^4$ in agreement with analytical results
\cite{Fischer:2002eq,Zwanziger:2001kw,Lerche:2002ep,Pawlowski:2003hq}.
This is a decisive difference, since it can be shown that an infrared vanishing 
gluon propagator cannot have a positive definite spectral function and is therefore
confined. Indeed, Zwanziger has argued that the lattice gluon propagator should vanish 
in the continuum limit \cite{Zwanziger:1991gz} and therefore be confined as an effect 
of the proximity of the Gribov-horizon for low momentum gauge field configurations.
However, no statement could be made as to the rate with which the continuum limit
behaviour is approached. Current extrapolations of lattice data to the infinite volume
limit (on large asymmetric lattices) are somewhat ambivalent. 
Whereas an extrapolation of the gluon dressing function leads to $\kappa \approx 0.52$
(IR-vanishing), one obtains $\kappa = 0.5$ (IR-finite)  from an extrapolation of the 
corresponding gluon propagator \cite{Silva:2005hb}. Therefore it seems as if there 
is a genuine difference between propagators on different manifolds, which has to be
taken into account when extrapolating to infinite volumes. The differences 
shown in Fig.~\ref{vol2} may serve as a  measure of the upper limit of these effects.

The (quenched) solution for the gluon propagator on both type of manifolds is used directly
as input in the corresponding quark-SDE on a torus and in the infinite volume/continuum limit. 
What remains then is to specify an explicit expression
for the quark-gluon vertex. Here we follow the strategy of Ref.~\cite{Bhagwat:2003vw}
and employ a parametrisation of the vertex such that the (quenched) lattice results for 
the quark propagator are reproduced by the gap equation. Our ansatz for the vertex is
\beq
\Gamma_\nu(k,\mu^2) \;=\; \gamma_\nu\, \Gamma_{1}(k^2)\, \Gamma_{2}(k^2,\mu^2)\, \Gamma_{3}(k^2,\mu^2) \label{v1}
\eeq
where $k^2$ is the gluon momentum and $\mu^2$ is the renormalisation scale. 
The ansatz depends on the gluon momentum only and is thus the simplest
possible form that respects charge conjugation symmetry. Furthermore this
choice of momentum dependence ensures the existence of a corresponding
kernel in the Bethe-Salpeter equation for mesons in accordance with the
axial-vector Ward-Takahashi identity, {\it cf.} section \ref{sec:pion}.
The three components of this ansatz are given by
\vspace{-3mm}
\beqa
\Gamma_{1}(k^2) &=&  \frac{\pi \gamma_m}{\ln(k^2/\Lambda_{QCD}^2 +\tau)}\,, \label{v2}\\[3mm]
\Gamma_{2}(k^2,\mu^2) &=& G(k^2,\mu^2)\ G(\zeta^2,\mu^2)\ \widetilde{Z}_3(\mu^2)\ h \ [\ln(k^2/\Lambda_{g}^2 +\tau)]^{1+\delta} \label{v3}\\[3mm]
\Gamma_{3}(k^2,\mu^2) &=& Z_2(\mu^2)\; \frac{a(M)+k^2/\Lambda_{QCD}^2}{1+k^2/\Lambda_{QCD}^2}\,, \label{v4}
\eeqa
where $\delta=-9/44$ is the (quenched) one-loop anomalous dimension of the ghost,
$\gamma_m=12/33$ the corresponding anomalous dimension of the quark and $\tau = e-1$ acts 
as a convenient infrared cutoff for the logarithms. It is well known that the effective interaction 
$g^2 Z \Gamma_{1} \Gamma_{2} \Gamma_{3}$ in the 
quark-SDE has to approach the running coupling in the ultraviolet momentum regime \cite{Miransky:1985ib}. 
In our ansatz this UV-part of the interaction is represented by $\Gamma_1$. 
The scale $\Lambda_{QCD}$ is scheme dependent. Here, since we fit to the
lattice data of Refs.~\cite{Bowman:2002bm,Zhang:2004gv}, its value
corresponds to the MOM-scheme used therein.
The product $Z \Gamma_{2} \Gamma_{3}$
goes to a constant for large momenta, since the ultraviolet behaviour of the ghost and gluon dressing functions 
is given by
\beq
G(z) \;=\; G(s)\left[\omega\log\left(\frac{z}{s}\right)+1\right]^\delta \,, \ \ \ 
Z(z) \;=\; Z(s)\left[\omega\log\left(\frac{z}{s}\right)+1\right]^\gamma \,,
\label{Z_UV}
\eeq
with $\omega=\beta_0\alpha(s)/(4 \pi)=11N_c\alpha(s)/(12 \pi)$ and a large scale $s$. The anomalous 
dimensions of the ghost and the gluon dressing functions in $Z \Gamma_{2} \Gamma_{3}$ combine to 
$\gamma+\delta=-1-\delta$, which is balanced by the explicit logarithm in $\Gamma_2$. 
The scale $\Lambda_{g}$ in Eq.~(\ref{v3}) represents a possibly scheme
dependent scale inherent in the Yang-Mills part of the interaction, which
is related to the analytic structure of the gluon propagator.
The coefficient $h$ is fixed such that the ultraviolet behaviour of the resulting running coupling matches
the one calculated on the lattice in Ref.~\cite{Sternbeck:2005tk}. The renormalisation group invariant 
$G(\zeta^2,\mu^2)\widetilde{Z}_3(\mu^2)$ with the arbitrary scale $\zeta$ is introduced to impose the correct
cutoff- and renormalisation point dependences of the effective interaction in the quark-SDE.
Together, the product $\Gamma_1 \Gamma_2$ represents the non-Abelian content of the quark-gluon
vertex as expressed in its Slavnov-Taylor identity (STI) given by \cite{Marciano:1978su}
\begin{equation}
G^{-1}(k^2) \: k_\nu \: \Gamma_\nu(q,k) = S^{-1}(p) \: H(q,p) - H(q,p) \: S^{-1}(q),
\label{quark-gluon-STI}
\end{equation}
where $q$ and $p$ are the quark momenta. This identity enforces the presence of the ghost factor
$G(k^2)$ in $\Gamma_2$, which makes the quark-gluon vertex an infrared singular object, similar to
the three- and four-gluon vertices \cite{Alkofer:2004it}.
The remaining part of $\Gamma_1 \Gamma_2$ is infrared finite and can be interpreted as a model
for the ghost-quark scattering kernel $H(q,p)$. 
The dependency of the vertex on the quark
wave function $Z_f$ through the inverse quark propagators in the STI is taken care of by $\Gamma_3$, the
form of which is chosen appropriately. 
The extra factor $Z_2$ is vital in ensuring
multiplicative renormalisability of the quark-SDE\footnote{Note that the effective
interaction of Ref.~\cite{Bhagwat:2003vw} fails to ensure multiplicative renormalisability and is therefore
only valid at the fixed renormalisation point chosen in their work.}. The dependence of this part of the
vertex on the quark mass is expressed in terms of the function
\beq
a(M)\; =\; \frac{a_1}{1 + a_2 M(\zeta^2)/\Lambda_{QCD} + a_3 M^2(\zeta^2)/\Lambda_{QCD}^2}.
\label{am}
\eeq
In order to preserve multiplicative renormalisability of the quark-SDE, it is important that
the scale $\zeta$ at which the quark mass function is read off (and also the ghost factor in $\Gamma_1$) is not 
correlated with the renormalisation point. Instead it should be a fixed scale
sufficiently far
into the ultraviolet region that volume effects are negligible. In our calculations we use $\zeta=2.9$ GeV.

\begin{table}
\hspace*{2cm}
\begin{center}
\begin{tabular}{|c||c|c|c||c|c|c|c|}\hline
&\multicolumn{3}{c||}{staggered} & \multicolumn{4}{c|}{overlap}                  \rule[-3mm]{0mm}{7mm}\\\hline\hline
$m_{lattice}$(GeV)       & \hspace*{1mm} 0.028 \hspace*{1mm} & \hspace*{1mm} 0.057 \hspace*{1mm} 
                         & \hspace*{1mm} 0.114 \hspace*{1mm} & \hspace*{1mm} 0.090 \hspace*{1mm} 
			 & \hspace*{1mm} 0.140 \hspace*{1mm} & \hspace*{1mm} 0.210 \hspace*{1mm}
			 & \hspace*{1mm} 0.300 \hspace*{1mm} \rule[-3mm]{0mm}{7mm}\\\hline
$M_{SDE}(\zeta^2)$(GeV)  & 0.044 & 0.080 & 0.151 & 0.076 & 0.112 & 0.162 & 0.225 \rule[-3mm]{0mm}{7mm}\\\hline
\end{tabular}
\vspace{3mm}
\caption{Renormalised quark masses in the SDE on the torus at $\zeta=2.9$ GeV compared with the bare quark
masses on the lattice from Ref.~\cite{Bowman:2002bm,Zhang:2004gv}. Note that the bare staggered quark masses
are smaller than the renormalised quark masses from the SDE, whereas the corresponding bare overlap
quark masses are larger (c.f. the discussion in the text).  \label{tab1}}
\vspace*{5mm}

\begin{tabular}{|c||c|c|c|c|c|c|}
\hline
          & \hspace*{2mm} h                    \hspace*{2mm} 
	  & \hspace*{1mm} $\Lambda_g$(GeV)     \hspace*{1mm} 
	  & \hspace*{1mm} $\Lambda_{QCD}$(GeV) \hspace*{1mm} 
	  & \hspace*{2mm} $a_1$                \hspace*{2mm}
	  & \hspace*{2mm} $a_2$                \hspace*{2mm} 
	  & \hspace*{2mm} $a_3$                \hspace*{2mm}\rule[-3mm]{0mm}{7mm}\\\hline\hline
staggered &  1.33 & 1.50  & 0.35 & 25.30 & 4.80  & -1.39 \rule[-3mm]{0mm}{7mm} \\ \hline
overlap   &  1.31 & 1.50  & 0.35 & 25.58 & 3.44  &  2.23 \rule[-3mm]{0mm}{7mm} \\\hline
\end{tabular}
\vspace{3mm}
\caption{Parameters used in the vertex model, Eqs.~(\ref{v1}-\ref{v4}).\label{tab2}}
\end{center}
\vspace{5mm}
\end{table}

To fit the various parameters in our model interaction we solve the quark-SDE on a torus employing a momentum 
range similar to that used in the lattice calculations of Ref.~\cite{Bowman:2002bm,Zhang:2004gv}. In the SDE this corresponds to
a $24^4$ ($36^4$) lattice in momentum space with a smallest momentum of $304$ MeV ($200$ MeV) corresponding to a box-size
of $L_1=L_2=L_3=L_4=L=2.04$ fm ($L=3.10$ fm) for the staggered (overlap) quarks. 
We first determine the parameters $h, \Lambda_g$ and $\Lambda_{QCD}$ by fitting
the 'Yang-Mills part' $\Gamma_1 \Gamma_2$ of our vertex model to the ultraviolet part of the
lattice running coupling from Ref.~\cite{Sternbeck:2005tk}, which agrees with the two-loop results
from perturbation theory. This fit is not unique but gives a range of pairs $(\Lambda_g,\Lambda_{QCD})$ related to the value of $h$. 
We then choose current quark masses $m(\mu^2)$ such that the ultraviolet behaviour of
the lattice quark mass functions are reproduced by the solutions of the quark SDE. As the ultraviolet behaviour 
of the quark mass function is determined by the ultraviolet behaviour of the input interaction, we find the same 
result for each pair $(\Lambda_g,\Lambda_{QCD})$. This merely reflects the fact that the 
ultraviolet behaviour of the quark mass function is controlled by resummed perturbation theory 
and is therefore model independent. We then deduce values of the function $a(m)$ in Eq.(\ref{am}) such that the lattice quark
mass functions are reproduced. Although this is possible for all values of the pairs $(\Lambda_g,\Lambda_{QCD})$,
the corresponding quark wave functions $Z_f(p^2)$ favour a small range of values for $\Lambda_{QCD}$ and
$\Lambda_g$. Finally we fix the coefficients $a_{1,2,3}$ and determine the values of the
quark mass function $M(\zeta)$ corresponding to different renormalised current quark masses $m(\mu^2)$. 
Our final best parameter sets together with the quark masses are tabulated in Tables \ref{tab1},\ref{tab2}. 
Note that the bare staggered quark masses are smaller than the renormalised quark masses from the SDE, 
whereas the corresponding bare overlap quark masses are larger. Comparing the functions $M_{SDE}(m_{lattice})$
for both formulations we find agreement if the bare staggered quark masses are multiplied by a factor $1.70$.
No additional additive corrections, which recently have been identified as a consequence of taste 
symmetry breaking \cite{Hasenfratz:2005ri}, are necessary. 

A few comments on the quality of our fits are in order.
The scales $\Lambda_{QCD}$ and $\Lambda_g$ cannot be determined very
well; we estimate the error in these
scales to be of the order of 30 \%. With the given values for these
scales, the uncertainty in the parameter $h$ is then only related to the
(small) error bars of the lattice coupling from
Ref.~\cite{Sternbeck:2005tk}, {\it i.e.} of the order of 2 \%. The errors
for our values of the parameters $a_1,a_2,a_3$ are correlated to
both the error in the determination of the current quark mass from the
lattice data and the error
bars of the lattice data in the infrared momentum regime. This leads to
uncertainties of the order of a few percent.
We have also performed fits to $a(M)$  setting $a_3=0$, thereby testing
the possible redundancy of this parameter. The fits
become worse in the range where lattice data are available, but still
might be of tolerable quality. However,
there is a nontrivial effect which convinced us of the importance of the
parameter $a_3$: if $a_3=0$ we obtained
vastly different values for $a(M)$ in the chiral limit for the staggered
and the overlap data. However, fitting
$a_3$ as a free parameter these values almost coincide. As a consequence,
 one obtains almost identical quark propagators
if the infinite volume/continuum limit and then the chiral limit are
performed. On general grounds the overlap and staggered formulation of
quarks on the lattice should coincide in the continuum limit. Possible
residual scheme dependences at finite current quark masses necessarily
vanish in the chiral limit, provided there are no additive mass
corrections (see above). This is exactly what we see if $a_3$ is included
as a free parameter. All qualitative conclusions of our paper do not
depend on the details of the fits and are stable
with respect to a variation of the parameters within the errors given
above. For our quantitative results we have
included these uncertainties in our estimate of the overall error
margins. Unfortunately it is extremely difficult to give a quantitative
estimate of the systematic truncation error, {\it i.e.} an estimate of
possible changes once subleading structures of the vertex are taken into
account. Wherever possible we have tried to assess
such effects on a qualitative basis.

\subsection{The quark propagator at finite and infinite volume}\label{sec:quark}

\begin{figure}[t]
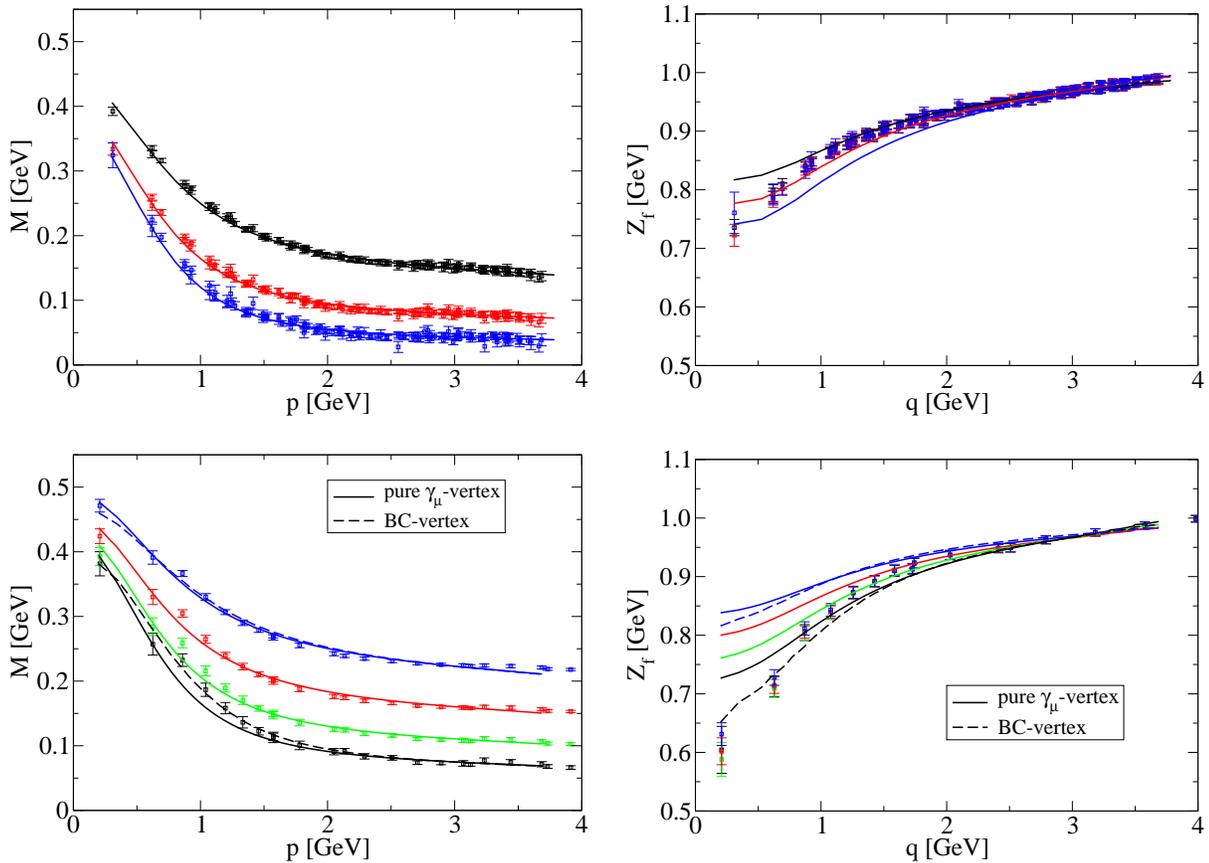

\centerline{
  \epsfig{file=torquark.fitM.eps,width=7.7cm}
  \hfill
  \epsfig{file=torquark.fitZ.eps,width=7.7cm}}
\vspace*{4mm}
\centerline{
  \epsfig{file=torquark.fitM.over.eps,width=7.7cm}
  \hfill
  \epsfig{file=torquark.fitZ.over.eps,width=7.7cm}}
\caption{The results for quark mass function $M$ and the wave function $Z_f$
from Dyson-Schwinger equations on the torus compared with lattice data for
staggered quarks \cite{Bowman:2002bm} (upper panel) and overlap quarks
\cite{Zhang:2004gv} (lower panel). The lattice momentum $p$ has been used for the 
lattice mass functions, whereas the choice of the \lq kinematical momentum' $q$ corrects for
hypercubical artifacts in the lattice wave functions \cite{Bowman:2002bm,Zhang:2004gv}.
The results for the Ball-Chiu ansatz, Eq.~(\ref{BC}), are shown only for the 
largest and smallest masses of the overlap quark propagator.   \label{fitMZ}}
\vspace*{5mm}
\end{figure}

\begin{figure}[t]
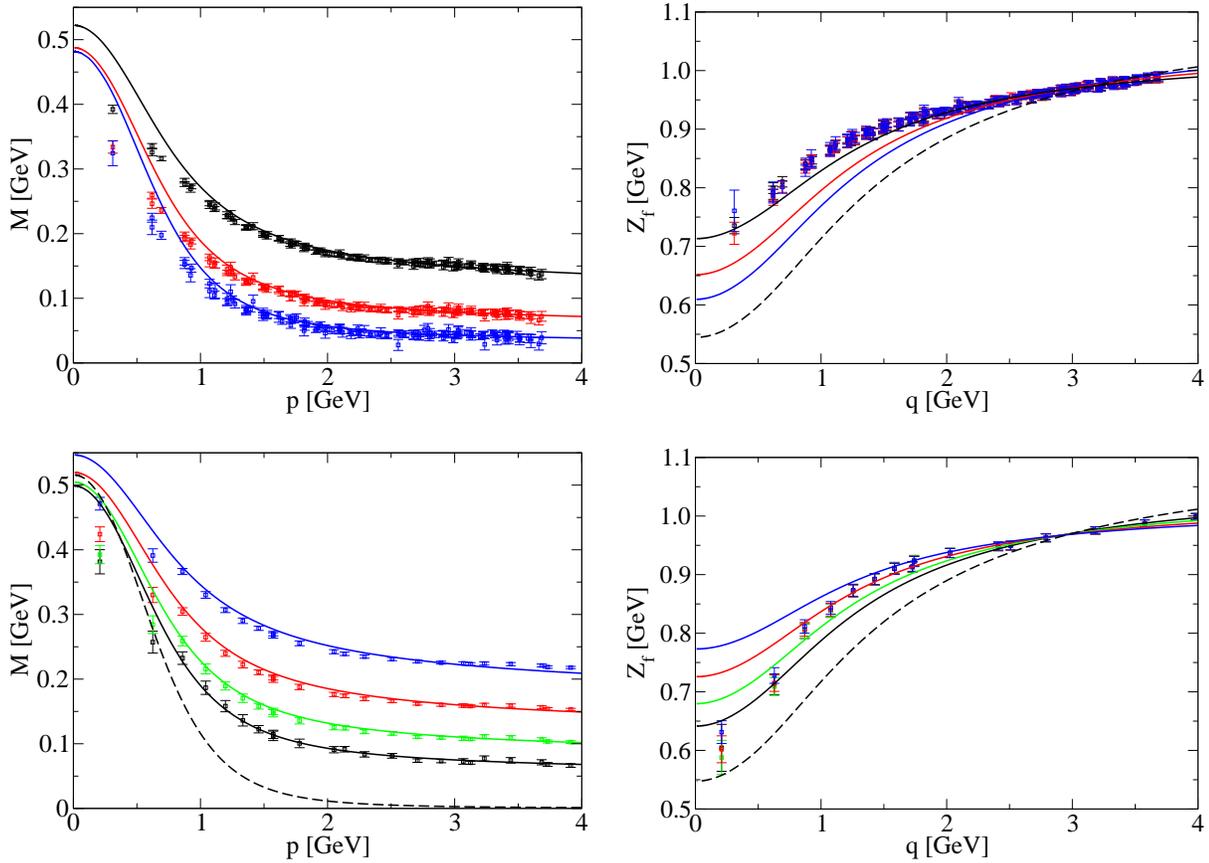

\centerline{
  \epsfig{file=torquark.contM.eps,width=7.7cm}
  \hfill
  \epsfig{file=torquark.contZ.eps,width=7.7cm}}
\vspace*{4mm}
\centerline{
  \epsfig{file=torquark.contM.over.eps,width=7.7cm}
  \hfill
  \epsfig{file=torquark.contZ.over.eps,width=7.7cm}}
\caption{The results for quark mass function $M$ and the wave function $Z_f$
from Dyson-Schwinger equations in the continuum compared with the same lattice data as in Fig.~4
(staggered quarks in the upper panel, overlap quarks in the lower panel).
Also shown is the chiral limit (dashed curves).\label{contMZ}}
\vspace{5mm}
\end{figure}

In Fig.~\ref{fitMZ} we compare our results from the SDE with the lattice data. The solutions for
the quark mass function $M(p^2)$ from the SDE can be nicely fitted to the lattice results in
both, the staggered and the overlap formulation. For the quark wave function $Z_f(p^2)$
in the staggered formulation, shown in the upper right diagram of Fig.~\ref{fitMZ}, 
we observe a slightly larger spread in the SDE solutions
than on the lattice\footnote{The parametrisation used in Ref.~\cite{Bhagwat:2003vw} leads to a 
similar large spread (not explicitly shown in their paper).}. The same is true when compared
to the overlap data, although here in addition we observe a larger fall in the infrared.
We have tried to reproduce this steeper decrease by considering various modifications
of our model interaction, Eqs.~(\ref{v1})-(\ref{v4}), but have not succeeded. In particular
introducing additional ghost factors, which correspond to an even more singular vertex
in the continuum, does not improve the situation. It seems to us, that
such a fall can only be reproduced when further tensor structures of the quark-gluon vertex is 
taken into account. Indeed, employing the Ball-Chiu construction \cite{Ball:1980ay}
\beqa
\nonumber
\Gamma^{BC}_\nu &=& \frac{1}{2}\, \left(A(p^2)+A(q^2)\right)\, \gamma_\nu\, + 
 \, \frac{1}{2}\,\left(A(p^2)-A(q^2)\right)\,\frac {(\pslash+\qslash)(p+q)_\nu}{p^2-q^2}\\[0.5mm]
&&\hspace{4.07cm}+\,i\,\left(B(p^2)-B(q^2)\right)\, \frac{(p+q)_\nu}{p^2-q^2} \label{BC}
\eeqa
instead of $\gamma_\nu$ (and modified parameters $a_{1,2,3}$), we could at least reproduce
the sharper decrease at small masses. However, as can be seen in Fig.~\ref{fitMZ}, the problem remains
for large masses. It is therefore not clear to us, whether the sharper 
low momentum decrease in the overlap data
should be taken seriously and really interpreted as an indication of the importance of a richer tensor structure
in the quark-gluon vertex, or whether one should prefer the staggered data and conclude that
the $\gamma_\nu$-part of the vertex is sufficient to reproduce the lattice results. We leave
this question open for future investigations in both the SDE-formalism and on the lattice,
and proceed using our simple construction, Eqs.~(\ref{v1})-(\ref{v4}). 

To assess finite volume effects we now compare the lattice/SDE-results on the compact manifold
with the infinite volume/continuum limit. To this end we solve the quark-SDE in the continuum
employing our lattice inspired ansatz for the quark-gluon vertex and the continuum solutions
for the ghost and gluon propagators, discussed in section \ref{sec:YM}, as input. This procedure
does take into account finite volume effects from three different sources: firstly from the gluon
propagator, secondly from the quark-gluon vertex via its ghost content in $\Gamma_{2}$ and thirdly 
effects generated by the quark-SDE itself. The parameters of the fitted quark-gluon interaction
are kept fixed, as are the renormalised current quark masses. In Fig.~\ref{contMZ} we compare our 
results in the infinite volume limit to the lattice data. We observe finite volume effects for 
momenta smaller than 1 GeV. The most pronounced effects occur in the infrared momentum region,
where the continuum mass functions approach finite values at vanishing momentum which are 
substantially larger than anticipated from the lattice data alone. We conclude that lattice data
underestimate the quark mass that is dynamically generated by a substantial amount.
The absolute size of this effect is approximately the same for all 
quark masses we have investigated, so the relative size becomes smaller for heavier quarks. 
Exactly as expected heavier quarks fit on finite volumes most readily, but here we have deduced 
the size of the volume effects for all masses.

\begin{figure}[t]
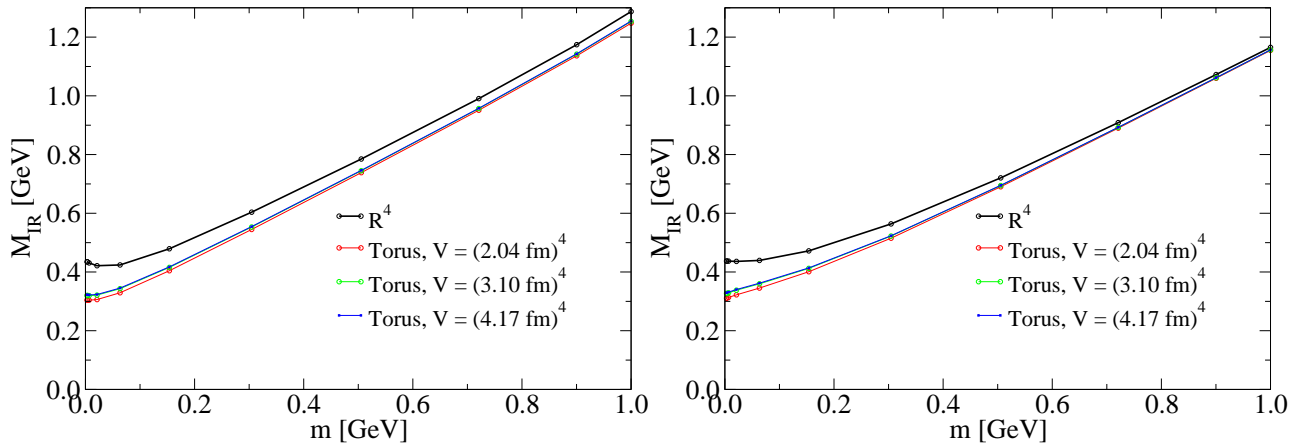

\centerline{
  \epsfig{file=torquark.Mm.eps,width=8.5cm}\hfill
  \epsfig{file=torquark.Mm.over.eps,width=8.5cm}}
\caption{The quark mass function at a given momentum 
$M_{IR} := M(p^2=0.0924 \mbox{GeV}^2)$ plotted as a function of the current quark
mass $m(\mu^2)$ in the continuum and on tori with different volumes
(staggered quarks in the left diagram, overlap quarks in the right diagram).
\label{Mm}}
\vspace{5mm}
\end{figure}

Another interesting effect occurs for small momenta:
the quark mass function in the chiral limit is actually larger than the one for the smallest
nonzero current quark mass. This effect has also been observed directly from the quenched
staggered lattice data \cite{Bowman:2005vx}. The effect is more pronounced for the staggered
quarks, as can be seen by comparing the two diagrams of Fig.~\ref{Mm} at small quark masses. 
There we show the quark mass function at a given momentum, $M_{IR} := M(p^2=0.0924 \mbox{~GeV}^2)$, 
plotted as a function of the current quark mass. Apart from small quantitative differences, 
the qualitative behaviour of $M_{IR}(m)$ is similar for both fermion formulations. 
On the compact manifold, one observes a (small)
volume dependence for $2 \, < L < 3$~fm, which becomes negligible for
$L > 3$~fm. However, even then one is still far away from the infinite volume results.
This remaining difference stems from the Yang-Mills sector of the theory: the difference
there between the torus and infinite volume results ({\it cf}. Fig.~\ref{vol2}) has a direct impact
on the effective interaction in the quark-SDE and therefore on the results for the quark
propagator. The size of this effect depends on the continuum value of the exponent $\kappa$ 
({\it cf}. eq.(\ref{Z_IR})), which is closely related to the quality of the approximation of the 
ghost-gluon-vertex in the ghost and gluon SDEs. There are indications that the \lq true'
exponent $\kappa$ may be closer to $\kappa \approx 0.5$ and therefore closer to the current 
lattice data than our value $\kappa \approx 0.596$ 
\cite{Lerche:2002ep,Pawlowski:2003hq,Zwanziger:2002ia}. Thus we regard our result as an upper limit 
for the effects that depend upon the volume.

\subsection{A critical volume for chiral perturbation theory}\label{sec:XPT}

\begin{figure}[t]
\centerline{
  \epsfig{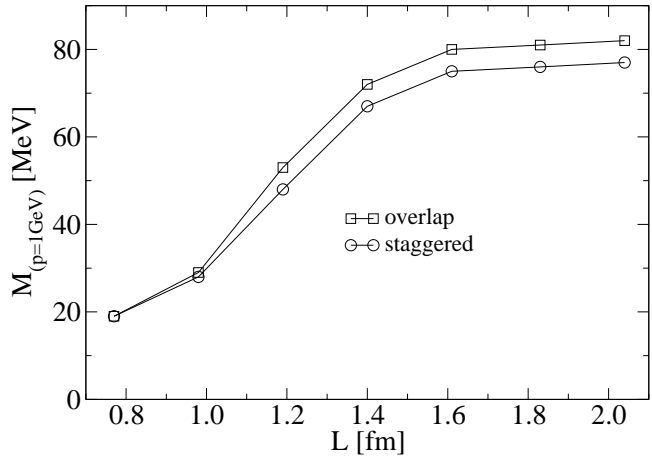}}
\caption{The quark mass function for an up/down quark at a given momentum $p=1\mbox{GeV}$ 
plotted as a function of the box length.  
\label{L}}
\vspace{3mm}
\end{figure}

As discussed above, dynamical chiral symmetry breaking alone cannot occur on a finite volume. A nonvanishing 
current quark mass satisfying the condition
\beq
V m |\langle \bar{q}q\rangle|\; >>\; \pi \label{LS}
\eeq
has to be present as a \lq seed' to trigger dynamical mass generation and the formation of a chiral condensate. 
Chiral perturbation theory, on the other hand, is built upon the chiral limit and therefore can only be applied on 
volumes large enough such that very small quark masses are feasible. The corresponding condition, discussed above, is
\beq
L\; >> \;\frac{1}{2 f_\pi} \sim 1 \,\,\mbox{fm}. \label{LS2}
\eeq
We will now give an estimate of how large $L$ has to be in practice. To this end we employ a current quark mass
of the order of a typical up/down-quark mass, 
$M(p^2=2.9 \mbox{GeV}^2)=10 \mbox{MeV}$
, and determine the mass function $M(p^2)$ at $p^2=1$ GeV from solutions 
on tori with different volumes by linearly interpolating on the corresponding momentum grids. The result is
plotted in Fig.~\ref{L}. We clearly see that the quark mass function grows rapidly in the range 
$1.0 \, <\, L \, <\,  1.6$~fm signalling the onset of dynamical chiral symmetry breaking. Above 
$L\,=\, 1.6$~fm, a plateau is reached. This picture does not change when we extract the mass function $M(p^2)$
at smaller momenta $p^2$ or when we employ even smaller quark masses.  Thus a safe value for $L$ should be at least  
\beq
L_{\chi PT}\; \simeq\; 1.6 \,\,\mbox{fm}.
\eeq
This is a surprisingly small value in the light of the condition Eq.~(\ref{LS}) and so provides some justification
for extending chiral perturbation theory to rather small volumes.

\section{Properties of quarks and pions in the infinite volume/continuum limit}\label{sec:cont}

\subsection{The chiral condensate}\label{sec:condensate}

Having studied the volume dependence of the quark propagator in some detail, we now focus on $R^4$ and investigate the
quark/meson sector employing our effective interaction which has been fixed by the lattice input. From the quark propagator 
$S_{\chi}$ in the chiral limit, we can determine the value of the (renormalisation point dependent) chiral condensate using
\beq
-\langle \bar{q}q\rangle (\mu^2) := Z_2(\mu^2) \, Z_m(\mu^2) \, N_c \,\mbox{tr}_D \int
\frac{d^4q}{(2\pi)^4} S_{\chi}(q,\mu^2) \,,
\label{ch-cond}
\eeq
where the trace is over Dirac indices. Its value is conveniently determined at a large 
renormalisation scale, converted to the renormalisation point independent chiral condensate 
and then run down to $\mu=2 $~GeV employing the quenched scale $\Lambda^{\overline{MS}}_{QCD}=0.225(21) \mbox{MeV}$ \cite{Capitani:1998mq}.
We then obtain the values 
\beq
-\langle \bar{q}q\rangle_{overlap}^{\overline{MS}} (\mu^2) \;=\; (252.6 \pm 5.0 \mbox{MeV})^3, \ \ \ 
-\langle \bar{q}q\rangle_{staggered}^{\overline{MS}} (\mu^2) \;=\; (253.0 \pm 5.0 \mbox{MeV})^3,
\eeq
which are in very good agreement with each other. The error given is an estimate of combined numerical and scale uncertainties.
It is a quite interesting and satisfying result that the continuum 
chiral limit quark propagators of the staggered and overlap fermions agree very well with each other. It is furthermore
interesting to compare our value for the chiral condensate with recent results using other methods on the lattice. 
Gimenez {\it et al} \cite{Gimenez:2005nt} find the value
$(265 \pm 27 \mbox{MeV})^3$ from an operator product expansion employing an $O(a)$-improved quenched Wilson 
action. Wennekers and Wittig \cite{Wennekers:2005wa} quote $(285 \pm 9 \mbox{MeV})^3$, determined
from a quenched overlap action. Both values are in fair agreement with each other and with our result. 
McNeile \cite{McNeile:2005pd} recently obtained the value $(259 \pm 27 \mbox{MeV})^3$ from a chiral Lagrangian with parameters fixed
by lattice data employing $N_f=2+1$ staggered sea quarks, thus indicating that unquenching effects in the
chiral condensate may be small. This is in excellent agreement with the prediction in the 
SDE/BSE-approach \cite{Fischer:2003rp,Fischer:2005en}.

\subsection{On the analytical properties of the quark propagator}\label{sec:analytical}

\begin{figure}[t]
\centerline{
  \epsfig{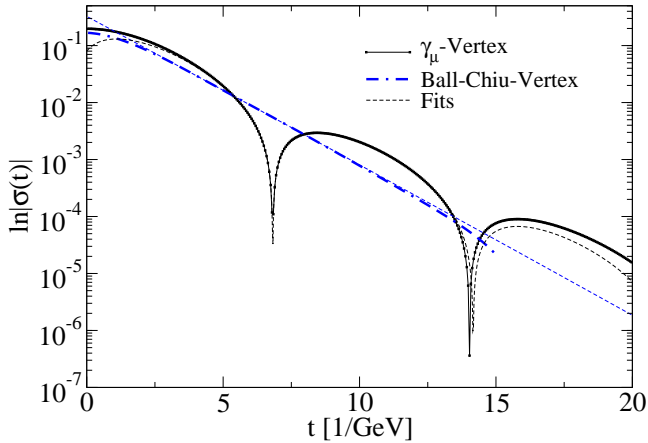}}
\caption{The logarithm of the Schwinger function $\ln(|\sigma(t)|)$ of the chiral
limit quark propagator as a function of time. Shown are results for the staggered 
and overlap quark, (i) employing the ansatz Eq.~(\ref{v1}) for the quark-gluon vertex 
and (ii) substituting the Ball-Chiu construction, Eq.(\ref{BC}), for $\gamma_\nu$.
The respective curves for the staggered and overlap quarks are indistinguishable 
in the plot. We compare the results to fits of the function Eq.~(\ref{cc}). 
\label{schwinger}}
\vspace{3mm}
\end{figure}

The analytic properties of the quark propagator can in part be read off from the corresponding
Schwinger function
\beq
\sigma(t)\; =\; \int d^3x \int \frac{d^4p}{(2\pi)^4} \exp(i p \cdot x) \sigma_{S,V}(p^2),
\eeq
where $\sigma_{S,V}$ are the scalar and the vector parts, respectively, of the dressed quark propagator.
(This method has a long history, see {\it e.g.} 
\cite{Atkinson:1978tk,Krein:1990sf,Burden:1991gd,Maris:1991cb,Oehme:1994pv,Burden:1997ja,Alkofer:2003jj} 
and references therein). According to the Osterwalder-Schrader axioms of 
Euclidean field theory \cite{Osterwalder:1973dx}, this function has to be positive 
to allow for asymptotic states 
in the physical sector 
of the state space of QCD. Conversely, positivity violations in the Schwinger function
show that the corresponding asymptotic states (if present) belong to the
unphysical part of the  state space. Thus positivity violations constitute a 
sufficient condition for confinement. Our results for the Schwinger function of the
chiral limit quark propagator in the staggered and overlap formalisms 
are shown in Fig.~\ref{schwinger}. The Schwinger functions of the two formulations
agree extremely well and are indistinguishable in the plot. Let us first discuss the result 
obtained with our simple vertex construction involving only the vector part $\gamma_\mu$ of 
the vertex. The cusp at $t=6.76$ GeV$^{-1}\,=\,1.33$ fm indicates a node in the
Schwinger function corresponding to positivity violations at a scale in rough agreement 
with the size of hadrons. An excellent fit to the Schwinger function is obtained using the form
\cite{Stingl:1996nk}
\beq
\sigma(t)\; =\; |\, b_0 \exp(-b_1 t) \cos(b_2 t+b_3)\,| \quad , \label{cc}
\eeq
which corresponds to a pair of complex conjugate poles of the propagator 
in the timelike momentum plane. These poles correspond to a \lq quark mass' given by
$m = b_1 \pm i b_2$, which in our case is  $m = 516(20) \pm i \, 428(20)$ MeV. Taken at face value
this means that the lattice quark propagator is confined (this is also the conclusion 
drawn by Bhagwat {\it et al.}~\cite{Bhagwat:2003vw}). However, there is a caveat: it has been shown in 
Ref.~\cite{Alkofer:2003jj} that the presence of a sufficiently strong
scalar part in the quark-gluon vertex can have a strong influence on the analytic
structure of the solution of the quark-SDE. Indeed if we employ the Ball-Chiu construction, Eq.(\ref{BC}), 
instead of $\gamma_\mu$ we obtain an exponentially decaying Schwinger function denoted by the
dash-dotted line in Fig.~\ref{schwinger}. Such a function corresponds to a positive definite
quark propagator with a pole on the real axis at $m = 632(20) \pm i \, 0(2)$ MeV 
(within numerical accuracy). This shows that the analytic structure of the quark
propagator depends strongly on the details of the structure of the quark-gluon vertex and one
cannot make definite statements from fitted interactions alone. First attempts on the lattice \cite{Skullerud:2003qu}
as well as in the SDE/BSE approach \cite{Bhagwat:2004hn,Bhagwat:2004kj,Fischer:2004ym}
have been made to study the tensor structure of the quark-gluon vertex
in more detail. Still, more effort is needed before definitive conclusions about the analytic structure of
the quark propagator are in sight.

\subsection{Pion mass and decay constant from the lattice interaction}\label{sec:pion}

\begin{table}
\begin{center}
\begin{tabular}{|c||c|c|c|}\hline
           & \hspace*{1mm} $m^{\mu= 2 GeV}_{\overline{MS}}$ (MeV) \hspace*{1mm}       
	   & \hspace*{1mm} $M_\pi$(MeV)                           \hspace*{1mm} 
	   & \hspace*{1mm} $f_\pi$(MeV)                           \hspace*{1mm}  \rule[-3mm]{0mm}{7mm}  \\\hline\hline
staggered  &  $4.1 \pm 0.3$  & 138.4        & 83.5         \rule[-3mm]{0mm}{7mm}   \\ \hline
overlap    &  $4.1 \pm 0.3$  & 138.7        & 84.3         \rule[-3mm]{0mm}{7mm}  \\\hline
experiment &                 & 138.5        & 92.4         \rule[-3mm]{0mm}{7mm}  \\\hline
\end{tabular}
\vspace{3mm}
\caption{Results for the renormalised current up/down quark mass, the mass of the pion
and the pion decay constant employing the quark-gluon interaction fitted to the lattice 
data. The evolution of the quark mass has been performed using 
$\Lambda^{\overline{MS}}_{QCD}=0.225(21) \mbox{MeV}$ \cite{Capitani:1998mq}.\label{pion}}
\end{center}
\vspace{3mm}
\end{table}

Finally we have determined the mass and the decay constant of the pion employing the 
effective quark-gluon 
interaction fitted to the lattice data. The pion is described by
the homogeneous Bethe-Salpeter equation (BSE)
\beq
\Gamma^\pi_{\alpha \beta}(p;P)\;=\;\int \frac{d^4k}{(2 \pi)^4}\, K_{\alpha \beta;\delta \gamma}(p,k;P)\,
\left[S(k_+)\Gamma^\pi(k;P)S(k_-)\right]_{\gamma \delta}
\label{eq:bse}
\eeq
where 
\beq
\Gamma^\pi(p;P) \;=\; \gamma_{5} \left[E^\pi(p;P)- \imath \Pslash \, F^\pi(p;P) \,
- \imath \pslash G^\pi(p;P)
- \left[\Pslash\,,\pslash \right]\, H^\pi(p;P) \right],\label{pseu}
\eeq
is the Bethe-Salpeter amplitude of the pion, $K$ 
is the Bethe-Salpeter kernel and the momenta $k_+=k+\xi P$ and $k_-=k+(\xi-1)P$ are 
such that the total momentum $P=k_+-k_-$. All physical results are 
independent of the momentum partitioning parameter $\xi=[0,1]$. The crucial link 
between the meson bound states and their quark and gluon constituents is provided by the 
axial vector Ward-Takahashi identity. It relates the quark self energy to the 
quark-quark interaction kernel in the BSE and thereby guarantees the Goldstone
nature of the pions and kaons \cite{Bender:1996bb,Maris:1997hd}. In our case the 
kernel is given by  
\beq
K_{\alpha \beta;\delta \gamma}(p,k;P)\rightarrow 
\left[\gamma_\mu \right]_{\alpha \gamma}
\left[\gamma_\nu \right]_{\delta \beta} 
t_{\mu\nu}(k) Z_2 \frac{Z(k) \, \Gamma_1(k)\,  \Gamma_2(k)\, \Gamma_3(k)}{k^2}\,,
\eeq
where $t_{\mu\nu}$ is a transverse projector in momentum space and the flavour 
content of the kernel has been suppressed. To determine the pion mass and wave functions
we explicitly solve the quark-SDE in the complex momentum plane, thereby providing
the necessary input for the BSE. The BSE is then solved as an eigenvalue equation for
eigenvalue one, which provides the wave functions and the pole mass of the pion 
(all technical details of such a calculation are discussed in detail in 
Ref.~\cite{Fischer:2005en}). From the (normalised) wave function the pion decay constant is fixed by
\beq
f_{\pi}\;=\;\frac{3}{M^2}\,{\rm Tr}_d\int \frac{d^4k}{(2 \pi)^4} \Gamma^\pi(k,-P)S(k+P/2)
\gamma_5 \Pslash (k-P/2).
\label{eq:fpi}
\eeq

Our results for the pion mass, the corresponding renormalised 
current quark mass in the $\overline{MS}$-scheme and the pion decay constant are given in Table \ref{pion}. 
The current quark mass has been determined from the quark-SDE at a large renormalisation point,
converted into the $\overline{MS}$-scheme and subsequently evolved to $\mu = 2 \,\mbox{GeV}$
employing the same scale as for the chiral condensate, 
$\Lambda^{\overline{MS}}_{QCD}=0.225(21) \mbox{MeV}$ \cite{Capitani:1998mq}.
The errors given in Table \ref{pion} are an estimate of numerical and scale uncertainties.
The resulting current quark mass is in the ballpark of the values quoted by
the particle data group \cite{Eidelman:2004wy}. 

Probably the most interesting observable is the pion decay constant, which directly
reflects the deficiency of the quenched lattice calculation compared with the
real world. Both lattice formulations underestimate the experimental value by roughly
ten percent. This margin is much smaller than the thirty percent estimated in 
Ref.~\cite{Bhagwat:2003vw}, where finite volume effects were not taken into account.  

\section{Summary}\label{sec:sum}

In summary, we have investigated the properties of a quenched lattice-QCD quark propagator 
from staggered quarks \cite{Bowman:2002bm} and from an overlap quark action \cite{Zhang:2004gv}
in Landau gauge. We employed a coupled set of Schwinger-Dyson equations for the ghost, gluon 
and quark propagators on compact manifolds and in the infinite volume/continuum limit to 
study finite volume effects in the propagators and to determine some aspects of dynamical 
chiral symmetry breaking on a torus. 
We constructed a model for the quark-gluon vertex such that two sets of staggered and overlap 
lattice quark propagators are reproduced by the quark gap equation on their respective manifolds. 

Comparing results on different volumes with the infinite volume/continuum limit, we found
sizeable quantitative but not qualitative differences at small momenta. The continuum mass 
functions approach finite values at vanishing momentum which are substantially larger than 
anticipated from the lattice data alone. Thus lattice simulations may underestimate 
the amount of dynamical quark mass generation in the infrared by as much as 100 MeV. 
The absolute size of this effect is approximately the same for all 
quark masses we have investigated, so the relative size becomes smaller for heavier quarks. 
As a by-product of this investigation we observed that the bare quark masses in the 
staggered and overlap formulations are related by a simple multiplicative factor. 
No additional additive corrections were necessary, which is a signal that taste symmetry 
violations in the staggered action used in Ref.~\cite{Bowman:2002bm} are negligible \cite{Hasenfratz:2005ri}.

We also assessed the effects of small volumes on dynamical chiral symmetry breaking. Employing
a fixed small current quark mass we decreased the volume of the box until we found clear
signals of chiral symmetry restoration. These signals occur at the surprisingly small box length
of 
\beq
L \simeq 1.6 \,\,\mbox{fm},
\eeq
which constitutes a minimal box size below which chiral perturbation 
theory cannot safely be applied.

With the quark-gluon interaction fixed by the lattice data, we then determined the properties of 
the quark propagator and pions in the infinite volume/continuum limit. We found a chiral condensate
of
\beq
|\langle \bar{q}q\rangle|_{\overline{MS}}^{2 GeV} = (253 \pm 5 \,\,\mbox{MeV})^3
\eeq
which compares favourably with values
determined with other methods on the lattice. Unfortunately nothing can be said about the
analytic structure of the quark propagator. We showed, that the method of fitting a quark-gluon
interaction to lattice propagator data is not sufficient to pin down the relative strength of the 
various tensor components of the vertex, which in turn are necessary to derive reliable 
statements on the analytic structure of the quark propagator. Finally we have determined the pion
mass and decay constant employing a rainbow-ladder truncation of the Bethe-Salpeter equation,
which incorporates the fitted interaction. We obtained the renormalised up/down quark masses
\beq
m_{\overline{MS}}^{2 GeV} = 4.1 \pm 0.3 \,\, \mbox{MeV},
\eeq
which are in the ballpark of the values given in the particle data book \cite{Eidelman:2004wy}.
The pion decay constant is roughly ten percent smaller than the experimental value. This indicates
that quenching effects in the light meson sector are not too large in agreement with previous 
findings \cite{Fischer:2005en}.

We have shown how the SDE/BSE approach, once matched to lattice results on finite volumes
with appropriate manifolds, can reliably determine infinite volume/continuum
predictions for all quark masses, including those of the real world close to the
chiral limit: a limit not directly accessible on the lattice.

\acknowledgments

We thank P. Bowman for providing us with the lattice data for the staggered quark 
and J. Zhang for the data of the overlap quark. We thank Jonivar Skullerud for
comments on the manuscript. The work of CSF has been supported by the 
Deutsche Forschungsgemeinschaft (DFG) under contract Fi 970/2-1.
MRP  acknowledges partial support of the EU-RTN Programme, 
Contract No. HPRN-CT-2002-00311, \lq\lq EURIDICE''.


\end{document}